\documentclass[RNAAS]{aastex62}

\newcommand\sfr{{\rm SFR}}

\newcommand\Mstel{M_{\ast}}

\usepackage{color}
\usepackage[]{hyperref}
\usepackage{mathptmx}
\usepackage{amsmath}

\begin{document}

\title{At Least One-in-Six Galaxies Is Always Dead}
%\title{Stochastic Theory Predicts a Time-Independent Null Passive Fraction of 10\%--20\%}

%% Note that the corresponding author command and emails has to come
%% before everything else. Also place all the emails in the \email
%% command instead of using multiple \email calls.
\correspondingauthor{L.E.~Abramson}
\email{labramson@carnegiescience.org}

%% The \author command can take an optional ORCID.
\author[0000-0002-8860-1032]{L.E.~Abramson}
\altaffiliation{Carnegie Fellow}
\affiliation{Carnegie Observatories, 813 Santa Barbara Street, Pasadena, CA 91101, USA}

\author[0000-0003-4727-4327]{D.D.~Kelson}
%\altaffiliation{RNAAS Editor}
\affiliation{Carnegie Observatories, 813 Santa Barbara Street, Pasadena, CA 91101, USA}

%% Note that RNAAS manuscripts DO NOT have abstracts.

\begin{abstract}
\noindent Via numerical experiments, we show that the $\sim$10\%--20\% passive fraction seen 
at $z\gtrsim3$ is consistent with galaxy star formation histories being maximally correlated stochastic processes. 
If so, this fraction should reflect a time-independent baseline that holds at any epoch or mass regime 
where mean star formation rates are rising. Data at $\log\Mstel\leq10$ and $z\lesssim0.5$ bear this 
out, as should future {\it James Webb Space Telescope} observations.
\end{abstract}


%% Start the main body of the article. If no sections in the 
%% research note leave the \section call blank to make the title.
\maketitle
%\section{} 

\vspace{1em} Galaxy star formation histories (SFHs) can be modeled as stochastic 
processes \citep{Kelson14,Kelson16,Kelson20}. In this framework, star formation rates (SFRs)
are non-negative sums of random numbers. The SFR at the epoch of observation, $T$, can
be independent of values at earlier times, $t$, or correlated with them, on average, via a 
weighting factor $(T-t)^{H-1/2}$. Here, $H$ defines the strength of the correlation 
\citep{Mandelbrot68}. If $H=1/2$, the SFH returns to a purely random Brownian walk. 
If $1/2<H\leq1$, past SFR changes bias future changes in the same direction. 

Ensembles of stochastic SFHs describe coeval galaxy populations. While the star formation rate
of any individual member cannot be predicted, the ensemble's mean and dispersion evolve 
deterministically in time as functions of $H$ \citep{Kelson14}:
\begin{align}
%	\langle\sfr(T)\rangle&=\sigma\,\sqrt{\frac{2}{\pi}}\,\frac{T^{H}}{2H}, \label{eq:mean}\\
	\langle\sfr(T)\rangle&\propto\frac{T^{H}}{H}, \label{eq:mean}\\
	\mathrm{Sig}[\sfr(T)]&=\sqrt{H}\,\langle\sfr(T)\rangle. \label{eq:var}
\end{align}
These equations have been shown to hold at redshifts preceding the peak of cosmic star formation
\citep{Kelson16} and in mass regimes where sSFRs ($\equiv\sfr/\Mstel$) are 
mass-independent \citep{Kelson14}. 

While an ensemble's mean SFR always rises in time, its dispersion grows more slowly for all $H<1$. 
So, at fixed epoch, more objects will reach $\sfr(T)=0$ if SFHs are maximally correlated ($H=1$) 
rather than random ($H=1/2$). Intuitively, this trend reflects the fact that random walks spend less 
time at $\sfr=0$ than ones in which past declines bias future changes toward zero. Physically, it 
means that the passive galaxy fraction constrains SFH stochasticity. 

\begin{figure}[h!]
\centering
\includegraphics[width = 6.5in, trim = 0cm 0cm 0cm 0cm]{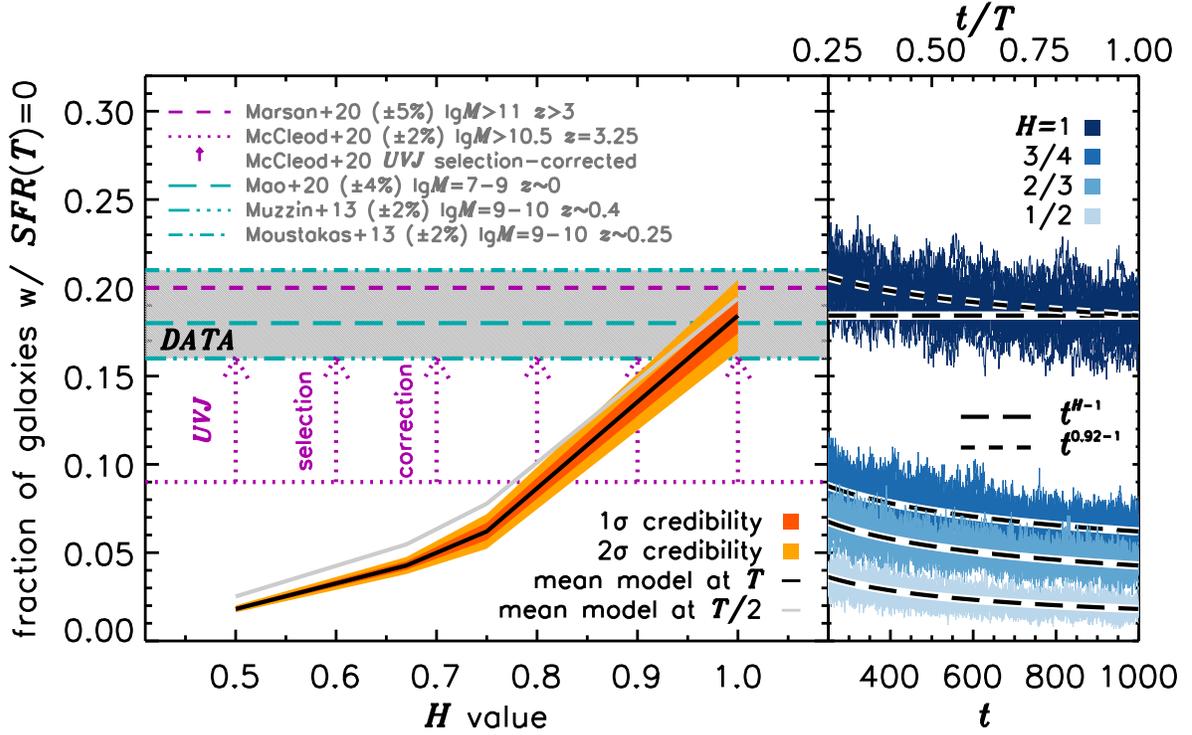}
\caption{{\it Left}: data at $z>3$ and $\log\Mstel<10$ at lower redshifts directly suggest that
	galaxy SFHs are consistent with $H\gtrsim0.8$ stochastic processes. Accounting for
	color 	selection effects, all data suggest $H\gtrsim0.9$, or maximal temporal correlation. The 
	data describe regimes in which fuel- (high-redshift) or quenching-related SFR constraints 
	(low-redshift) are absent. {\it Right}: if $H\simeq1$---as the SFR--$\Mstel$ relation and 
	stellar mass function shape also imply---then a null passive fraction of 16\%--20\%
	should be constant in time (uppermost long dashes). At all $H<1$,
	the passive fraction declines as $T^{H-1}$ contrary to observations. 
	We start the plot at $t/T=0.25$ as numerical convergence issues affect the early 
	timesteps as $H\rightarrow1$. Our $H_{\rm input}=1$ may thus reflect 
	$H_{\rm eff}\simeq0.92$, but the implied passive fraction evolves so slowly---staying 
	above 15\% after 4,000 timesteps---that the results at {\it left} should capture the true 
	null passive fraction. (See also grey line denoting fractions at $T/2$ at {\it left}.)}
\label{fig:frac}
\end{figure}

Figure \ref{fig:frac}, left, illustrates the above trend in 20 numerical ensembles of 1000 SFHs 
spanning 1000 timesteps. At $H=1/2$, only 2\% of objects have $\sfr=0$ in the last timestep. By $H=1$,
18\% do. The overplotted measurements of the passive fraction at $z\gtrsim3$ and 
$\log\Mstel\lesssim10$ at $z\lesssim0.5$---where the stochastic framework most directly 
applies---clearly favor $H\gtrsim0.8$. 

\citet{McLeod20}'s $\sim$9\% implies the lowest $H$, but simulations show 
that these authors' {\it UVJ} passive selection---based on \citet[][]{Williams09}'s $z<0.5$ 
prescription---may miss 45\% of intrinsically non-starforming 
galaxies at $z\sim3$.\footnote[2]{Tests assumed a Gaussian $A_{V}\leq2$\,mag distribution with a 
mean and dispersion of 0.5\,mag, $\log Z/Z_{\odot}=-0.5$, 
a \citet{Chabrier03} initial mass function, and no photometric errors. Using the relevant observed bands 
yields better results at $z=3$ ({\it F140W}, $K_{s}$, {\it Spitzer} IRAC2).} 
If so, those data imply a true passive fraction of $\sim$16\%, suggesting $H\simeq1$. Such selection 
effects worsen with increasing redshift: by $z\sim4$, only about $1/3$ of passive galaxies are accurately 
identified. If so, \citet{Santini20}'s $z\gtrsim4$ estimate of a $\sim$7\% passive fraction
($\log\Mstel\geq10.5$) is consistent with \citet{Marsan20}'s $z>3$ measurement 
of 15\%--25\% ($\log\Mstel\geq11$) and also suggests $H\simeq1$,\footnote{Technically, the spectral 
templates \citet{Santini20} use to identify passive galaxies come from a library based on a 
non-evovled {\it UVJ} cut such that $1/2$ to $2/3$ of truly passive spectral energy 
distributions may go unmatched.} as the lower-redshift data clearly 
prefer \citep{Muzzin13,Moustakas13,Mao20}.

%z	Np	Ntot        Np,rec %p,rec %inUVJ
%1.0ÊÊÊ1756ÊÊ10000ÊÊ1464 0.8337 0.1762
%2.0ÊÊÊ1756ÊÊ10000ÊÊ1290 0.7346 0.1524
%3.0ÊÊÊ1756ÊÊ10000ÊÊÊ975 0.5552 0.1129
%4.0ÊÊÊ1756ÊÊ10000ÊÊÊ510 0.2904 0.0571

Measuring the passive fraction is difficult, but we are confident in the above statements because results 
based on the SFR--stellar mass relation \citep{Kelson14}, stellar mass functions \citep{Kelson16}, and 
the growth of cosmic structure \citep{Kelson20} all independently imply that $H\simeq1$. Intriguingly, 
so does an analytical derivation based on Newtonian gravitational collapse and mass conservation 
\citep{Kelson20}. $H\simeq1$ also fits with our intuition that galaxies' observed properties should 
meaningfully reflect their pasts \citep{Caplar19,Iyer20}. 

Indeed, the fact that the same passive fraction describes epochs as disparate as those spanned by the 
above datasets suggests $H\simeq1$ on its own. The portion of tracks near the minimum of a stochastic 
process ($\sfr=0$) is analytically known to evolve as $T^{H-1}$ \citep[][]{Molchan00}. This 
trend---confirmed numerically in Figure \ref{fig:frac}, right---is incompatible with all observations, 
confining $H$ near unity a priori. If so, the passive fraction should be time-independent as is observed.

Additional processes separate what we have calculated---the fraction of galaxies expected
to have zero star formation given normal fluctuations in hospitable conditions---from the ``quenched'' 
fraction measured at higher masses or later times \citep[e.g.,][]{Peng10}. The framework discussed 
here does not yet capture those forces, so our results should be interpreted as a floor or null 
expectation for the passive fraction in complete ensembles.

As such, we predict that future unbiased JWST surveys will find 16\% to 20\% of galaxies to 
be classifiable as passive at all $z\gtrsim3$, and also at lower redshifts in mass regimes where 
sSFRs are mass-independent. We look forward to forthcoming tests!

%% An example figure call using \includegraphics

\clearpage


\begin{thebibliography}{}
\expandafter\ifx\csname natexlab\endcsname\relax\def\natexlab#1{#1}\fi

\bibitem[{{Caplar} \& {Tacchella}(2019)}]{Caplar19}
{Caplar}, N., \& {Tacchella}, S. 2019, \mnras, 487, 3845

\bibitem[Chabrier(2003)]{Chabrier03} Chabrier, G.\ 2003, \pasp, 115, 763

\bibitem[Iyer et al.(2020)]{Iyer20} Iyer, K.~G., Tacchella, S., Genel, S., et al.\ 2020, \mnras, 498, 430

\bibitem[{{Kelson}(2014)}]{Kelson14}
{Kelson}, D.~D. 2014, ArXiv e-prints, arXiv:1406.5191

\bibitem[{{Kelson} {et~al.}(2016){Kelson}, {Benson}, \& {Abramson}}]{Kelson16}
{Kelson}, D.~D., {Benson}, A.~J., \& {Abramson}, L.~E. 2016, ArXiv e-prints,
  arXiv:1610.06566

\bibitem[Kelson et al.(2020)]{Kelson20} Kelson, D.~D., Abramson, L.~E., Benson, A.~J., et al.\ 2020, \mnras, 494, 2628%. doi:10.1093/mnras/staa100

\bibitem[{{Mandelbrot} \& {Van Ness}(1968)}]{Mandelbrot68}
{Mandelbrot}, B.~B., \& {Van Ness}, J.~W. 1968, SIAM Review, 10, 422

\bibitem[Mao et al.(2020)]{Mao20} Mao, Y.-Y., Geha, M., Wechsler, R.~H., et al.\ 2020, arXiv:2008.12783

\bibitem[Marsan et al.(2020)]{Marsan20} Marsan, Z.~C., Muzzin, A., Marchesini, D., et al.\ 2020, arXiv:2010.04725

\bibitem[McLeod et al.(2020)]{McLeod20} McLeod, D.~J., McLure, R.~J., Dunlop, J.~S., et al.\ 2020, arXiv:2009.03176

\bibitem[Molchan(2000)]{Molchan00} Molchan, G.~M. 2000, Theory Probab.~Appl., 44(1), 97--102.

\bibitem[Moustakas et al.(2013)]{Moustakas13} Moustakas, J., Coil, A.~L., Aird, J., et al.\ 2013, \apj, 767, 50%. doi:10.1088/0004-637X/767/1/50

\bibitem[Muzzin et al.(2013)]{Muzzin13} Muzzin, A., Marchesini, D., Stefanon, M., et al.\ 2013, \apj, 777, 18%. doi:10.1088/0004-637X/777/1/18

\bibitem[Peng et al.(2010)]{Peng10} Peng, Y.-. jie ., Lilly, S.~J., Kova{\v{c}}, K., et al.\ 2010, \apj, 721, 193%. doi:10.1088/0004-637X/721/1/193

%\bibitem[Romano \& Wolf (2000)]{Romano00}Romano, J.~P., \& Wolf, M.~2000, Statistics \& Probability Letters
%47, 115

\bibitem[Santini et al.(2020)]{Santini20} Santini, P., Castellano, M., Merlin, E., et al.\ 2020, arXiv:2011.10584

\bibitem[Williams et al.(2009)]{Williams09} Williams, R.~J., Quadri, R.~F., Franx, M., et al.\ 2009, \apj, 691, 1879

\end{thebibliography}
\end{document}